\documentclass[twocolumn,showpacs,preprintnumbers]{revtex4}%
\usepackage{amsmath}
\usepackage{amsfonts}
\usepackage{graphicx}
\usepackage{amssymb}%
\providecommand{\U}[1]{\protect\rule{.1in}{.1in}}
\providecommand{\U}[1]{\protect\rule{.1in}{.1in}}
\begin{document}
\title{Matter-wave bistability in coupled atom-molecule quantum gases}
\author{Lei Jiang$^{1}$, Han Pu$^{1}$ Andrew Robertson$^{2,3}$, and Hong Y. Ling$^{2}$}
\affiliation{$^{1}$Department of Physics and Astronomy, and Rice
Quantum Institute, Rice University, Houston, TX 77251, USA }
 \affiliation{$^{2}$Department of Physics and
Astronomy, Rowan University, Glassboro, New Jersey, 08028-1700, USA}
\affiliation{$^{3}$ Joint Quantum Institute and Condensed Matter
Theory Center, Department of Physics, University of Maryland,
College Park, MD 20742-4111, USA}

\begin{abstract}
We study the matter-wave bistability in coupled atom-molecule quantum gases,
in which heteronuclear molecules are created via an interspecies Feshbach
resonance involving either two-species Bose or two-species Fermi atoms at zero
temperature. We show that the resonant two-channel Bose model is equivalent to
the nondegenerate parametric down-conversion in quantum optics, while the
corresponding Fermi model can be mapped to a quantum optics model that
describes a single-mode laser field interacting with an ensemble of
inhomogeneously broadened two-level atoms. \ Using these analogy and the fact
that both models are subject to the Kerr nonlinearity due to the two-body
s-wave collisions, we show that under proper conditions, \ the population in
the molecular state in both models can be made to change with the Feshbach
detuning in a bistable fashion.

\end{abstract}

 \maketitle

\section{Introduction}

We study the matter-wave bistability in coupled atom-molecule quantum gases,
in which heteronuclear molecules are created via an interspecies Feshbach
resonance involving either two-species Bose or two-species Fermi atoms at zero
temperature. \ We show that the resonant two-channel Bose model is equivalent
to the nondegenerate parametric down-conversion in quantum optics, while the
corresponding Fermi model can be mapped to a quantum optics model that
describes a single-mode laser field interacting with an ensemble of
inhomogeneously broadened two-level atoms. \ Using these analogies and the
fact that both models are subject to the Kerr nonlinearity due to the two-body
s-wave collisions, we show that under proper conditions, \ the population in
the molecular state in both models can be made to change with the Feshbach
detuning in a bistable fashion.

\bigskip

The ability to cool and trap neutral atoms down to quantum degenerate regime
has created a host of new and exciting problems that are increasingly
interdisciplinary, bridging in particular the atomic, molecular, and optical
physics and the condensed matter physics. The rich knowledge and experience
accumulated over the past several decades in these fields have dramatically
accelerated the progress of ultracold atomic physics. An example that serves
to illustrate how the interdisciplinary fields learn and benefit from each
other is the phenomonon of atomic pairing where a bosoinc molecule is coupled
to two bosonic or fermionic constituent atoms via Feshbach resonance or
photoassociation. So far this is the only viable approach to create ultracold
molecules. It is also an ideal test ground for studying coupled atom-molecule
condensates and the BCS-BEC crossover \cite{levin}. The latter is thought to
be underlying the mechanism of high temperature superconductors and
extensively studied in the realm of condensed matter physics. In addition, the
coupled atom-molecule systems have deep quantum optical analogies
\cite{jack,Tikhonenkov}: bosoinc molecules coupled to bosonic atoms (which we
will refer to as the bosonic model in this paper) is the matter-wave analog of
parametric coupling of photons which has important applications in generating
nonclassical light fields and, more recently, in quantum information science;
while the system of bosonic molecules coupled to fermionic atoms (which we
will refer to as the fermionic model) can be mapped to the Dicke model where a
light field interacts with an ensemble of two-level atoms, a model having
fundamental importance in the field of quantum optics.

In this work, we will further explore these quantum optical analogies of the
atom-molecule system and focus on the important effects of binary collisional
interactions between atoms which are largely ignored in previous studies
\cite{jack,Tikhonenkov}. We show that the atom-atom interaction introduces
extra nonlinear terms which, under certain conditions, give rise to
matter-wave bistability in both bosonic and fermionic models. Hence, we may
establish the connection between the coupled atom-molecule quantum gases and
the nonlinear bistable systems \cite{gibbs} that have been extensively studied
in the 80's in the context of nonlinear optics, due both to its fundamental
interest, and to its many practical applications in fast optical switches,
optical memory, laser pulse shaping, etc.

\section{Bosonic model}

In what we call the bosonic model, a molecule associated with annihilation
operator $\hat{a}_{m}$ is coupled to two non-identical atoms labeled as
$|\uparrow\rangle$ and $|\downarrow\rangle$ with corresponding annihilation
operators $\hat{a}_{\uparrow}$ and $\hat{a}_{\downarrow}$, respectively. Here
we consider two types of atoms in order to make direct comparisons with the
fermionic model to be treated in the next section, for which only unlike
fermionic atoms can pair with each other and form a bosonic molecule.
Futhermore, in this work we only consider zero-temperature homogeneous case so
that all the bosons are condensed into zero center-of-mass momentum states.

The second quantized Hamiltonian reads
\begin{equation}
\hat{H}=\delta\,\hat{a}_{m}^{\dagger}\hat{a}_{m}+g\left(  \hat{a}_{m}%
^{\dagger}\hat{a}_{\uparrow}\hat{a}_{\downarrow}+h.c.\right)  +\sum_{i,j}%
\chi_{ij}\hat{a}_{i}^{\dagger}\hat{a}_{j}^{\dagger}\hat{a}_{j}\hat{a}_{i}\,,
\label{eq7}%
\end{equation}
where the detuning $\delta$ represents the energy difference between the
molecular and atomic levels which can be tuned by external field, $g$ is the
atom-molecule coupling strength and $\chi_{ij}=\chi_{ji}$ is the $s$-wave
collisional strength between modes $i$ and $j$. This system has been studied
in Ref.~\cite{zhoulu}. For completeness and better comparison with the
fermionic model, we briefly state some of the main results relevant to the
focus of this work --- matter-wave bistability --- and direct readers to
Ref.~\cite{zhoulu} for more details.

For our purpose, we take the standard mean-field approximation and replace
operators $\hat{a}_{j}$ with $c$-numbers $a_{j}=\sqrt{N_{j}}\,e^{i\varphi_{j}%
}$. The mean-field Hamiltonian takes the form:
\begin{equation}
H=2\Lambda(y^{2}-y)+2\nu y+(1-2y)\sqrt{2y}\,\cos\varphi\,, \label{ch}%
\end{equation}
where
\[
y=0.5\left[  1-\left(  N_{\uparrow}+N_{\downarrow}\right)  /N\right]
=N_{m}/N\,,\;\;\;\varphi=\varphi_{\uparrow}+\varphi_{\downarrow}-\varphi
_{m}\,,
\]
are a pair of conjugate variables, representing the molecular population and
phase mismatch, respectively. \ Other quantities are defined as
\begin{align*}
G  &  =g\sqrt{2N}\,,\\
\Lambda &  =N\left(  \chi_{\uparrow\uparrow}+\chi_{\downarrow\downarrow}%
+\chi_{mm}+2\chi_{\uparrow\downarrow}-2\chi_{m\uparrow}-2\chi_{m\downarrow
}\right)  /G\,,\\
\nu &  =\left[  \delta+\chi_{\uparrow\uparrow}+\chi_{\downarrow\downarrow
}+\left(  N-1\right)  \chi_{mm}-N\chi_{m\uparrow}-N\chi_{m\downarrow}\right]
/G\,,
\end{align*}
with $N\equiv N_{\uparrow}+N_{\downarrow}+2N_{m}$ a constant of motion
representing the total number of atoms, and we have assumed that the number of
atoms in states $|\uparrow\rangle$ and $|\downarrow\rangle$ are equal, i.e.,
$N_{\uparrow}=N_{\downarrow}$. In addition, we will focus on the stationary
states with $\varphi=\pi$ which has lower energies than the ones with
$\varphi=0$.

\subsection{Quantum Optical Analogy}

It is quite clear from the form of the second-quantized Hamiltonian in
Eq.~(\ref{eq7}) that without the collisional terms our model will reduce to
the trilinear Hamiltonian describing the nondegenerate parametric
down-conversion in quantum optics \cite{djb93,ma98}. In this analogy, the
molecular mode plays the role of the pump photon, where the two atomic modes
are the signal and idler photons, respectively. The collisional terms would
correspond to the Kerr-type cubic nonlinearity which will be present in the
optical system if the light fields propagate in some nonlinear medium
\cite{shen84}.

\subsection{Bistability}

\label{bb} 
\begin{figure}[ptb]
\begin{center}
\includegraphics[
width=2.5in
]{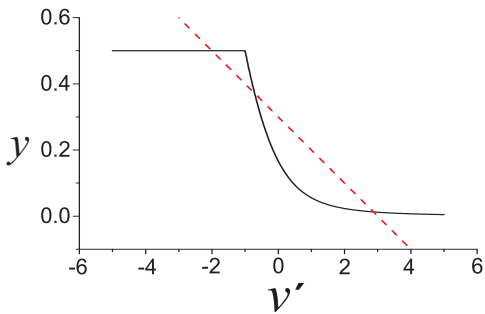}
\end{center}
\caption{For given $\Lambda$ and $\nu$, the thick solid line represents
$y_{0}(\nu^{\prime})$ and the thin dashed straight line represents
Eq.~(\ref{nu}). The intersects are the stationary solutions. Here we take
$1/2\Lambda=-0.1$ and $\nu=0.4\Lambda$.}
\label{Boson}%
\end{figure}

In the absence of the collisions or Kerr nonlinearity (i.e.,
$\Lambda=0$), the system does not exhibit bistability. This can be seen by
studying the properties of the mean-field Hamiltonian $H$ in Eq.~(\ref{ch})
which can be simplified as (taking $\varphi=\pi$)
\begin{equation}
H=2\nu y-(1-2y)\sqrt{2y}\,, \label{bh}%
\end{equation}
The stationary state correspond to the solution of
\begin{equation}
\frac{\partial H}{\partial y}=2\nu+3\sqrt{2y}-1/\sqrt{2y}=0\,. \label{sta1}%
\end{equation}
For a given detuning $\nu$, the stationary state is unique:
\begin{equation}
y_{0}(\nu)=\left\{
\begin{array}
[c]{ll}%
0.5\,, & \nu<-1\\
\frac{1}{18}(-\nu+\sqrt{\nu^{2}+3})^{2}\,, & \nu\geq-1
\end{array}
\right.  \label{bh1}%
\end{equation}
For $\Lambda\neq0$, using Eq.~(\ref{ch}), the stationary condition is given
by
\begin{equation}
\frac{\partial H}{\partial y}=2\nu^{\prime}+3\sqrt{2y}-1/\sqrt{2y}=0\,,
\label{sta2}%
\end{equation}
where we have defined
\begin{equation}
\nu^{\prime}=\nu+\Lambda(2y-1)\,. \label{nu}%
\end{equation}
Note that Eqs.~(\ref{sta1}) and (\ref{sta2}) have the same form. In other
words,we can express the effect of collisions as a nonlinear phase shift for
molecules that modifies the detuning $\nu$. Consequently, the stationary
solution for $\Lambda\neq0$ should have the same form as in Eq.~(\ref{bh1})
but with $\nu$ replaced by $\nu^{\prime}$, which makes $y_{0}$ an implicit
function of the detuning $\nu$. To find the explicit dependence of $y_{0}$ on
$\nu$, we can use the graphic method as illustrated in Fig.~\ref{Boson}. For
the example given, we obtain three stationary states. Further analysis shows
that the middle solution is dynamically unstable and the other two are stable
solutions \cite{zhoulu}. Such a behavior is typical in bistable systems
\cite{gibbs}.

The graphics of Fig.~\ref{Boson} also shows that, in order to have
multiple stationary solutions, the slope of the straight line (given
by $1/2\Lambda$) must be negative and cannot be too steep. More
specifically, the slope of the straight line has to be larger than
the slope of the curve at $\nu=-1$, and this leads to the condition
\begin{equation}
\Lambda<-1\,,\label{condition 1}%
\end{equation}
in order for the system to exhibit bistability.

\section{Fermionic model}

In the fermionic model, we denote $\hat{a}_{\mathbf{k},\sigma}$ as the
annihilation operator for an atom with spin $\sigma\left(  =\uparrow
\text{,}\downarrow\right)  $, momentum $\hbar\mathbf{k}$, and energy
$\epsilon_{k}=\hbar^{2}k^{2}/(2m)$, and as before denote $\hat{a}_{m}$ as the
annihilation operator for a molecule in state $\left\vert m\right\rangle $
with zero momentum. the second quantized Hamiltonian reads:
\begin{align}
\hat{H} &  =\sum_{\mathbf{k},\sigma}\epsilon_{k}\hat{a}_{\mathbf{k}\sigma
}^{\dag}\hat{a}_{\mathbf{k}\sigma}+U\sum_{\mathbf{k},\mathbf{k}^{\prime
},\mathbf{q}}\hat{a}_{\mathbf{k}\uparrow}^{\dag}\hat{a}_{-\mathbf{k}%
+\mathbf{q}\downarrow}^{\dag}\hat{a}_{-\mathbf{k}^{\prime}+\mathbf{q}%
\downarrow}\hat{a}_{\mathbf{k}^{\prime}\uparrow}\nonumber\\
&  +\nu\hat{a}_{m}^{\dag}\hat{a}_{m}+\frac{g}{\sqrt{V}}\sum_{\mathbf{k}%
}\left(  \hat{a}_{m}^{\dag}\hat{a}_{-\mathbf{k}\downarrow}\hat{a}%
_{\mathbf{k}\uparrow}+h.c.\right)  \,,\label{H1}%
\end{align}
where $V$ is the quantization volume. Hamiltonian (\ref{H1}) has the
form of the two-channel model of BCS-BEC crossover where only the
condensed molecule part is considered \cite{twochannel}. Following
the Hartree-Fock-Bogoliubov mean-field approach \cite{gennes89} by
dividing the two-body collision into a part related to the BCS gap
potential $\Delta=Up,$ where
\[
p=\sum_{\mathbf{k}}\left\langle \hat{a}_{-\mathbf{k}\downarrow}\hat
{a}_{\mathbf{k}\uparrow}\right\rangle /V\,,
\]
and a part related to the Hartree potential
\begin{equation}
V_{h}=U\sum_{\mathbf{k\sigma}}\langle\hat{a}_{\mathbf{k}\sigma}^{\dag}\hat
{a}_{\mathbf{k}\sigma}\rangle/(2V)\,,
\label{VH}
\end{equation}
where we again assume equal population in $|\uparrow\rangle$ and
$|\downarrow\rangle$ atomic states, i.e., $\langle\hat{a}_{\mathbf{k}\uparrow
}^{\dag}\hat{a}_{\mathbf{k}\uparrow}\rangle=\langle\hat{a}_{\mathbf{k}%
\downarrow}^{\dag}\hat{a}_{\mathbf{k}\downarrow}\rangle$, we may express the
Hamiltonian as%
\begin{align}
\hat{H} &  =\sum_{\mathbf{k},\sigma}(\epsilon_{k}+V_{h})\hat{a}_{\mathbf{k}%
\sigma}^{\dag}\hat{a}_{\mathbf{k}\sigma}+\nu\hat{a}_{m}^{\dag}\hat{a}%
_{m}\nonumber\\
&  +\sum_{\mathbf{k}}\left[  \left(  Up+{g}\hat{a}_{m}/\sqrt{V}\right)
\,\hat{a}_{\mathbf{k}\uparrow}^{\dag}\hat{a}_{-\mathbf{k}\downarrow}^{\dag
}+h.c\right]  \,.\label{H}%
\end{align}
Defining $\hat{N}=2\hat{a}_{m}^{\dag}\hat{a}_{m}+\sum_{\mathbf{k},\sigma}%
\hat{a}_{\mathbf{k}\sigma}^{\dag}\hat{a}_{\mathbf{k}\sigma}$ as the number
operator which is a constant of motion, we may rewrite the term proportional
to $V_h$ in (\ref{H}) as
\begin{align}
\sum_{\mathbf{k},\sigma}V_{h}\hat{a}_{\mathbf{k}\sigma}^{\dag}\hat
{a}_{\mathbf{k}\sigma} &  =V_{h}(\hat{N}-2\hat{b}^{\dag}\hat{b})\nonumber\\
&  =V_{h}\hat{N}-\left(  Un-{2U}\langle\hat{b}^{\dag}\hat{b}\rangle/V\right)
\hat{b}^{\dag}\hat{b}\,,\label{har}%
\end{align}
where $n=\langle\hat{N}\rangle/V$ is the constant total atom number
density. In our derivation, $V_h$ arises from the two-body atom-atom
collision. In general, additional terms representing atom-molecule
and molecule-molecule collisions are also present. These additional
terms will modify the coefficient $U$ in the definition of $V_h$
[Eq.~(\ref{VH})], which is the counterpart of $\Lambda$ in the
bosonic model, but the general form of Eq.~(\ref{har}) will remain
valid. In the following, we will refer to this term as the
collisional term. Through Eq.~(\ref{har}), we have expressed the
effect of the two-body collisions as a nonlinear energy shift of the
molecules (along with a constant energy bias $V_{h}N$), in complete
analogy with the bosonic model. We remark that in the usual
one-channel model of the mean-field BCS theory valid when the
molecular population is negligible, the collisional term just
represents an unimportant constant energy shift.

As usual, $\hat{a}_{\mathbf{k}\sigma}\left(  t\right)  $ and $\hat{a}%
_{m}\left(  t\right)  $ obey the Heisenberg equations of motion based on
Hamiltonian (\ref{H}). By replacing Bose operator $\hat{a}_{m} $ with the
related c-number $c=\langle\hat{b}\rangle/\sqrt{V}$ and Fermi operators
$\hat{a}_{\mathbf{k}\sigma}\left(  t\right)  $ with the familiar $u_{k}\left(
t\right)  $ and $v_{k}\left(  t\right)  $ parameters through the Bogoliubov
transformation $\hat{a}_{\mathbf{k}\uparrow}=u_{k}^{\ast}\,\hat{\alpha
}_{\mathbf{k}\uparrow}+v_{k}\,\hat{\alpha}_{-\mathbf{k}\downarrow}^{\dag}$ and
$\hat{a}_{-\mathbf{k}\downarrow}^{\dag}=-v_{k}^{\ast}\,\hat{\alpha
}_{\mathbf{k}\uparrow}+u_{k}\,\hat{\alpha}_{-\mathbf{k}\downarrow}^{\dag}$,
where $\hat{\alpha}_{\mathbf{k}\sigma}$ are the Fermi quasiparticle operators,
we arrive at the following set of mean-field equations of motion
\begin{subequations}
\label{dyneq}%
\begin{align}
i\hbar\dot{c}  &  = \nu_{e}c+gp\,,\label{dCm/dt}\\
i\hbar\dot{u}_{k}  &  = -\epsilon_{k} u_{k} + \Delta_{e} v_{k} \,,\label{uk}\\
i\hbar\dot{v}_{k}  &  = \Delta_{e} u_{k}+\epsilon_{k} v_{k} \,, \label{vk}%
\end{align}
where $p=\sum_{\mathbf{k}}u_{k}^{\ast}v_{k}/V$, $\Delta_{e}=gc+Up\,$, and
\end{subequations}
\begin{equation}
\nu_{e}=\nu-Un+2U\left\vert c\right\vert ^{2}\,, \label{nue}%
\end{equation}
is the effective detuning which contains a Kerr nonlinear term $2U|c|^{2}$
whose origin can be traced to the two-body collisional shift. This set of
equations describes the dynamics at zero temperature where the state of the
system can be described as the quasiparticle vacuum.

\subsection{Quantum Optical Analog}

\begin{figure}[ptb]
\begin{center}
\includegraphics[
width=2.5in
]{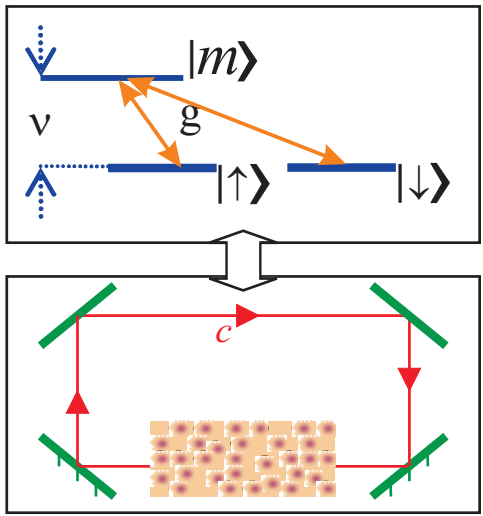}
\end{center}
\caption{(Color online) Mapping of the two-channel resonant Fermi superfluid
model to the Dicke model. The bosonic molecules and the fermionic atoms in the
former are mapped to the cavity laser field and an ensemble of two-level atoms
in the latter, respectively. See text for details.}%
\label{Fig:Model}%
\end{figure}

In several previous studies where the collisional term is neglected, it has been pointed out that the fermionic model can
be mapped to the Dicke model in quantum optics \cite{search,jack} as
schematically shown in Fig.~\ref{Fig:Model} (see below for details).
In fact, this model was recently shown to display collective dynamics similar
to photon echo and soliton-like oscillations in transient collective coherent
optics \cite{soliton}. Such a connection can be traced to the work of
Anderson's spin analogy \cite{anderson} for the BCS problem.

To show what is the quantum optical analogy of the collisional term,
let us rewrite Eqs.~(\ref{dyneq}) in a form more familiar in cavity optics. To
this end, we first introduce a set of new variables \[P_{k}=2u_{k}^{\ast}v_{k}%
\,,\;\;\;D_{k}=\left\vert u_{k}\right\vert ^{2}-\left\vert v_{k}\right\vert ^{2}\,,
\;\;\; \mathcal{E}_{L}=2i\Delta_{e}\,,\] and recast Eqs.~(\ref{uk}), (\ref{vk})
into
\begin{subequations}
\label{medium}%
\begin{align}
\hbar\dot{P}_{k}  &  =-i2\epsilon_{k}P_{k}-\mathcal{E}_{L}D_{k}\,,\label{Pk}\\
\hbar\dot{D}_{k}  &  =\left(  \mathcal{E}_{L}^{\ast}P_{k}+\mathcal{E}_{L}%
P_{k}^{\ast}\right)  /2\,. \label{Dk}%
\end{align}
Interpreting $P_{k}$ and $D_{k}$ as the microscopic polarization and
population inversion, respectively, Eqs.~(\ref{medium}) then become the
optical Bloch equation that describes the interaction between a local
electromagnetic field $\mathcal{E}_{L}$ and a fictitious two-level atom,
characterized with a transition energy $2\epsilon_{k}$ \cite{eberly}. This
analogy is consistent with the fact that there exists a one-to-one mapping
between pairs of fermion operators and Pauli matrices when the BCS pairing
mechanism is taken into account \cite{anderson}.

In this optical analogy, the local electric field $\mathcal{E}_{L}%
=\mathcal{E}+\mathcal{E}_{i}$ contains two contributions because of
$\Delta_{e}=gc+Up$. The first of these ($\mathcal{E}=i2gc$) is equivalent to
an average macroscopic field, whose dynamics is described by Eq.~(\ref{dCm/dt}%
), which can now be interpreted as the Maxwell's equation for the
cavity field $\mathcal{E}$ with cavity detuning $\nu_{e}$,
driven by a macroscopic polarization density $p=\sum_{\mathbf{k}}P_{k}/(2V)$
of an inhomogeneously-broadened medium [see Fig.~\ref{Fig:Model}].
The second part $\mathcal{E}_{i}=i2Up$ may be regarded as the internal field
at the atom due to the collective dipole polarization of the nearby two-level
atoms in the ensemble. As such, $\mathcal{E}_{L}=\mathcal{E}+\mathcal{E}_{i}$
here bears a direct analogy to the Lorentz-Lorenz relation in optics
\cite{jackson}. Note that had the collisional term been neglected (i.e., $U=0$),
there would have been no internal field contribution, nor would there have
been the Kerr nonlinearity in the equation for the bosonic mode. For $U\neq0$,
both of these terms will be present. Under such a circumstance,
Eqs.~(\ref{dCm/dt}) and (\ref{medium}) represent the generalized optical-Bloch
equations in which the Lorentz-Lorenz relation is explicitly incorporated
\cite{internal}, and hence can lead to interesting nonlinear phenomena just as
they do in optical systems.

\subsection{Bistability}

Having established this analogy, we now look for the steady state solution
from Eqs.~(\ref{dCm/dt}) and (\ref{medium}). As is well-known, the operation
frequency of a laser field is not known \textit{a priori}; but is established
through the so-called mode pulling --- the dynamical competition between
atomic and cavity resonances. A similar argument holds for the molecular field
$c$. For this reason, we adopt the following steady-state ansatz
\end{subequations}
\[
c\rightarrow c\,e^{-2i\mu t/\hbar},\;P_{k}\rightarrow P_{k}\,e^{-2i\mu
t/\hbar},\;D_{k}\rightarrow D_{k}%
\]
where the same symbols are used for both dynamical and steady-state variables
for notational simplicity. The molecular chemical potential, 2$\mu$, is just
the corresponding lasing frequency in the cavity optics model. From the steady
state equations obtained by inserting this stationary ansatz into
Eqs.~(\ref{dCm/dt}) and (\ref{medium}), we can easily find that (a) there
always exists a trivial solution or a \textquotedblleft
non-lasing\textquotedblright\ state with $\Delta_{e}=0$ or equivalently $c=0$,
which corresponds to the non-superfluid normal Fermi sea; and (b) a
non-trivial solution with its $\mu$, $\Delta_{e}$ and $c$
determined\textbf{\ }self-consistently from the gap equation%
\begin{equation}
\frac{1}{U-g^{2}/(\nu_{e}-2\mu)}=-\frac{1}{2V}\sum_{\mathbf{k}}\frac{1}{E_{k}%
}\,, \label{gap}%
\end{equation}
with $E_{k}=\sqrt{(\epsilon_{k}-\mu)^{2}+\Delta_{e}^{2}}$, the number
equation
\begin{equation}
2|c|^{2}+\frac{1}{V}\sum_{\mathbf{k}}\left(  1-\frac{\epsilon_{k}-\mu}{E_{k}%
}\right)  =n, \label{density}%
\end{equation}
and an auxiliary relation
\begin{equation}
|g\Delta_{e}|=|c(\nu_{e}-2\mu)[U-g^{2}/(\nu_{e}-2\mu)]|\,. \label{auxiliary}%
\end{equation}
The integral in the gap equation (\ref{gap}) under the assumption of contact
interaction is known to be ultraviolet divergent. To eliminate this problem,
we renormalize the interaction strength $U$ and $g$, as well as the detuning
$\nu$ in (\ref{gap}), while $U$ in the collisional term is replaced by the
background interaction strength $U_{0}$
\cite{renormalization,stoof}.

Note that there exists, in the single-mode inhomogeneously broadened laser
theory \cite{laser}, a similar set of steady-state integral equations, which,
due to lasers being open systems, are obtained under different considerations.
For example, the requirement that the cavity loss balance the saturated gain
leads to the \textquotedblleft gap\textquotedblright\ equation, whose primary
role is to limit the laser intensity; while the phase matching condition
translates into the \textquotedblleft number\textquotedblright\ equation,
whose main responsibility is to assign the amount of mode pulling of the laser
field relative to the cavity resonance.
\begin{figure}
[ptbh]
\begin{center}
\includegraphics[
height=1.1449in,
width=3.1963in
]%
{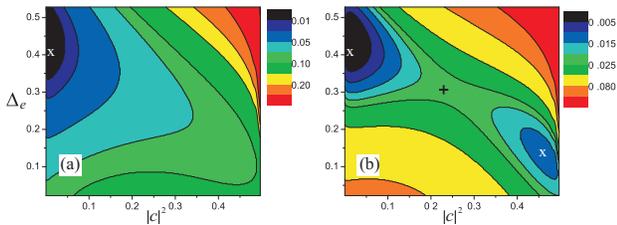}%
\caption{(Color online) Free energy density $f$ as a function of $\Delta_{e}$
and $|c|^{2}$ at $\nu=0.02$ (a) and $\nu=0.2$ (b). Extremum points are
indicated by `x' (minimum) and `+' (saddle point). $f$, $\Delta_{e}$, and
$\nu$ are all in units of $E_{F}=(3\pi^{2}n)^{2/3}/(2m)$, the Fermi energy of
the non-interaction system. In all the examples shown in this paper, the
physical parameters corresponding to $g_{0}$ and $U_{0}$ are $1.2\,E_{F}%
/k_{F}^{3/2}$ and $-60.7\,E_{F}/k_{F}^{3}$, respectively.}%
\label{free1}%
\end{center}
\end{figure}

An alternative way to derive Eqs.~(\ref{gap})-(\ref{auxiliary}) is from the
energy density. The zero-temperature energy density $f(\Delta_{e},c,\mu
)\equiv\langle\hat{H}\rangle/V$ can be calculated using Hamiltonian (\ref{H})
and the Bogoliubov transformation as \cite{stoof}
\begin{equation}
f=\sum_{\mathbf{k}}\frac{\epsilon_{k}-\mu-E_{k}}{V}-\frac{(\Delta_{e}-gc)^{2}%
}{U}+(\nu_{e}-2\mu)|c|^{2}+\mu n\,. \label{free}%
\end{equation}
The extremum conditions $\partial f/\partial\Delta_{e}=\partial f/\partial
c=0$, lead to Eqs.~(\ref{gap}) and (\ref{auxiliary}), respectively, while the
condition $\partial f/\partial\mu=0$ results in the number equation
(\ref{density}).

Figure \ref{free1} illustrates the energy density in the $|c|^{2}$-$\Delta
_{e}$ plane for different detuning $\nu$. For any given pair of ($c$,
$\Delta_{e}$), $\mu$ is calculated self-consistently using the number equation
(\ref{density}). Typically, $f$ has only one extremum which
is a minimum point as shown in Fig.~\ref{free1}(a). However, in the regime
$\nu\in(-0.08,0.13)\,E_{F}$, $f$ possesses three extrema:
two of them are local minima and the third a saddle point. An example with
$\nu=0.02$ is shown in Fig.~\ref{free1}(b).

\begin{figure}
[ptb]
\begin{center}
\includegraphics[
width=3in
]%
{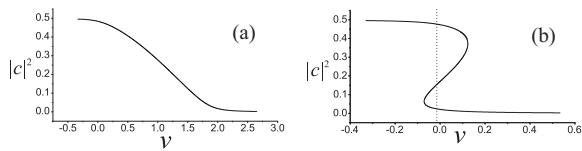}%
\caption{Molecular population $|c|^{2}$ as a function of detuning.
Vertical line in (a) indicate the critical point of a first-order phase
transition. In (a) the collisional term is included while it is neglected in (b).}%
\label{steady}%
\end{center}
\end{figure}

To gain more insights into the bistable behavior, we may carry an analogous
analysis as in Sec.~\ref{bb}. In the absence of the collisional term, steady-state
molecular population $|c|^{2}$ is a smooth monotonically decreasing function
of $\nu$ and the system does not exhibit bistability: As $\nu$ increases,
molecules decompose into atoms. This is shown in Fig.~\ref{steady}(a). When
collisional term is included, the relevant equations of motion maintain the same forms if we
substitute $\nu$ by
\begin{equation}
\nu^{\prime}=\nu+2U_{0}|c|^{2}\,. \label{nuf}%
\end{equation}
Hence the solution $|c|^{2}$ as a function of $\nu^{\prime}$ is represented by
the same curve as in Fig.~\ref{steady}(a). To find $|c|^{2}$ as a function of
$\nu$, we need to find the intersects between this curve and the straight line
representing Eq.~(\ref{nuf}). In direct analogy to the graphic method
in Fig.~\ref{Boson}, for $U_{0}$ sufficiently large and negative, these two
curves have three intersects and the system exhibits bistability. One example
is shown in Fig.~\ref{steady}(b). The vertical line in Fig.~\ref{steady}(b)
indicate the critical point of a first-order phase transition: across this
line, the ground state jumps from the upper branch to the lower one. For the
parameters used, this occurs at $\nu_{c}=-0.01\,E_{F}$.

To check the stability of these steady states, we have solved the
dynamical equations (\ref{dyneq}) using the slightly perturbed
steady state solution as the initial condition. From the dynamical
evolution of the system one can see that, just like in the bosonic
model, the states in the upper and lower branches are dynamically
stable: when slightly perturbed, they exhibit damped oscillations
around their equilibrium values. These oscillations can be further
understood from the excitation spectrum of the corresponding steady
state. This can be done using a linear stability analysis, which is
also the standard tool for studying laser instabilities
\cite{laser,lasers}. The spectrum is found to contain a discrete
part which determines the oscillation frequencies, and a continuous
part which contributes to the damping of these oscillations at a
much longer time scale \cite{CPT}. By contrast, the states in the
middle branch are unstable as small perturbations will lead to large
departures.


\subsection{Dynamics}

\begin{figure}[ptb]
\begin{center}
\includegraphics[
width=2.4151in
]{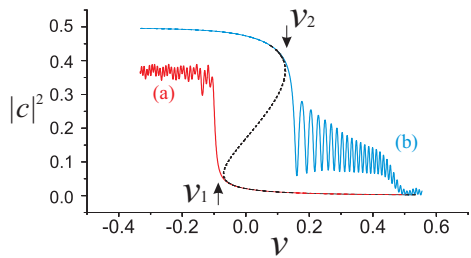}
\end{center}
\caption{(Color online) Dynamics of atom-molecule conversion as illustrated by
the molecular population when the detuning $\nu$ is slowly swept. Curve (a) is
obtained by sweeping $\nu$ from positive to negative values, while curve (b)
is obtained by sweeping $\nu$ in the opposite direction. The dotted line is
the steady-state molecular population, the same as in Fig.~\ref{steady}(b).}%
\label{sweep}%
\end{figure}

The bistability has important ramifications in atom-molecule conversion
dynamics. When the collisional term is unimportant and negligible, one can
easily create bosonic molecules from fermionic atoms by adiabatically sweeping
the Feshbach detuning across the resonance. As long as the sweeping speed is
sufficiently slow, the molecular population will follow the steady-state curve
as shown in Fig.~\ref{steady}(a). By contrast, when bistability induced by
the collisional term occurs, the adiabaticity condition will necessarily break
down. Fig.~\ref{sweep} displays the dynamical evolution of the bosonic
population when the detuning is swept starting either from a large positive or
a large negative value. We can see that the steady-state curve can be followed
up to the point where the stable states of the upper and lower branches and
the unstable states of the middle branch join each other (indicated by
$\nu_{1}$ and $\nu_{2}$ in Fig.~\ref{sweep}), where the population suddenly
jumps between the two stable branches. Note that the critical detuning
$\nu_{c}$ for the first-order phase transition as indicated by the vertical
line in Fig.~\ref{steady}(b) lies between $\nu_{1}$ and $\nu_{2}$. The
dynamical population curve thus exhibits hysteresis in the vicinity of the
first-order phase transition. In this way, by tuning the detuning in the
vicinity of $\nu_{1}$ or $\nu_{2}$, an atom-molecule switch can be realized.
Similar behavior is also found in the bosonic model.

\section{Conclusion}

In conclusion, we have studied the matter-wave bistability in coupled
atom-molecule quantum gases in both the bosonic and the fermionic models.
These two cases can be mapped to two very different quantum optical models:
parametric downconversion in the former and generalized Dicke model in the
latter. Nevertheless, one important common feature for both cases is that
bistability can be induced by collisional interactions which give rise to Kerr
nonlinearity. We hope that our work will motivate experimental efforts in
demonstrating the matter-wave bistability we predicted here.

\acknowledgments We thank Dr. Satyan Bhongale for useful discussions. HP
acknowledges support from NSF, the Robert A. Welch Foundation (Grant No.
C-1669), and the W. M. Keck Foundation, and HYL acknowledges support from NSF
and ARO.

\end{document}